\documentclass[11pt,twoside]{article}
\usepackage{newpasp}
\usepackage{newpasp,epsf}
\def\edcomment#1{\iffalse\marginpar{\raggedright\sl#1\/}\else\relax\fi}
\reversemarginpar

\begin{document}
\title{A multi-frequency study of the radio galaxy NGC326}
\author{M. Murgia}
\affil{Dip. Astronomia, Univ. di Bologna, 
via Ranzani 1, 40127 Bologna, Italy}
\affil{Ist. di Radioastronomia del CNR, via Gobetti 101, 40129 Bologna, Italy}

\author{P. Parma}
\affil{Ist. di Radioastronomia del CNR, via Gobetti 101, 40129 Bologna, Italy}

\author{R. Fanti}
\affil{Dip. Fisica, Univ. di Bologna, 
via Irnerio 46, 40127 Bologna, Italy}

\author{H. R. De Ruiter}
\affil{Oss. Astronomico di Bologna, 
via Ranzani 1, 40127 Bologna, Italy}

\author{R. D. Ekers}
\affil{ATNF, CSIRO, P.O. Box 76,
 Epping, NSW 2121, Australia}

\author{E. B. Fomalont}
\affil{NRAO, 520 Edgemont Road, Charlottesville, VA 2293, USA}

\begin{abstract}
We present preliminary results of a multi-frequency study of the inversion
 symmetric radio galaxy NGC326 based on VLA observations at 1.4, 1.6, 4.8, 8.5, and 14.9 GHz. These data allow us to investigate in detail the morphological, 
spectral and polarization properties of this peculiar object at different 
levels of spatial resolution.
\end{abstract}
\section{Introduction}
NGC326 belongs to the significant fraction of radio galaxies that are 
characterized by a distorted morphology, but also show some degree of symmetry.
 Such sources can by classified into two broad schemes: the mirror symmetric
 (`C'-shapes) and inversion symmetric (`Z'-shapes). There exists a sub-class 
of `Z'-symmetry, called `X'-shapes, which has two separated relic 
wings almost perpendicular to the currently active lobes (e.g. 3C315). 
There is a general consensus about the dynamical interpretation of the 
`C'-shapes. This kind of distortion is either caused by the translational 
motion of the jets with respect to the intergalactic medium (wide and narrow 
angle tails) or by the orbital motion of the galaxy around a nearby
 companion (e.g. 3C31). The explanation of the `Z' and `X'-shapes is still
 uncertain and matter of debate. The models proposed in the literature
 invoke jet precession (Rees 1978, Ekers et al. 1978), jet realignment
 (Wirth et al. 1982), and the buoyancy of the material in the lobes 
(Worrall et al. 1995).
\section{Source morphology} 
\begin{figure}
\plotfiddle{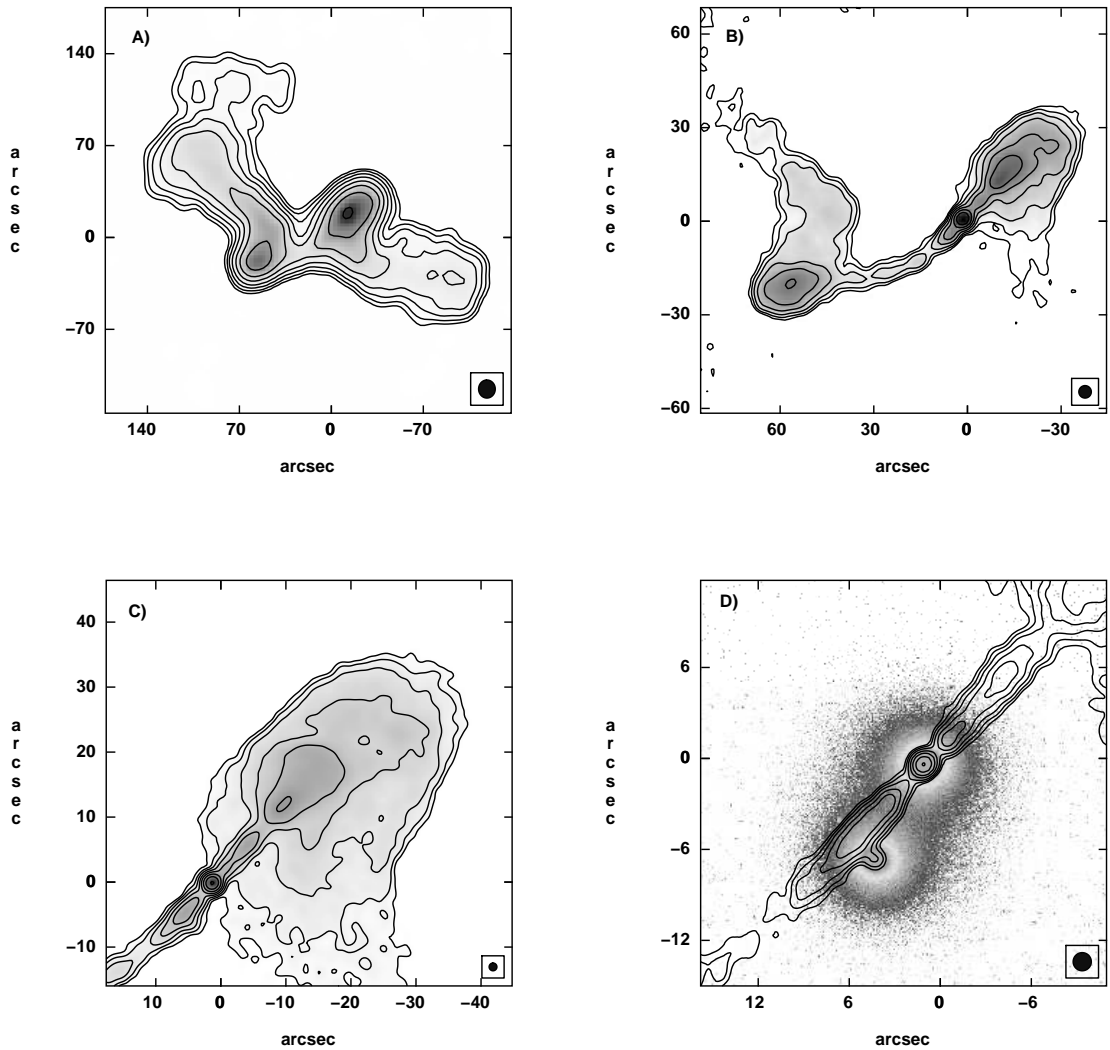}{16cm}{0.0}{80}{80}{-225.0}{-100.0}
\caption{A) VLA A+C array 1.4 GHz, the HPBW is 10\arcsec$\times$10\arcsec.
 B) VLA C+D array 14.9 GHz, the HPBW is 4\arcsec$\times$4\arcsec . 
C) VLA B+C array 4.8 GHz, the HPBW is 2\arcsec$\times$2\arcsec.
 D) VLA C array 1.4 GHz 2\arcsec$\times$2\arcsec resolution image of NGC326
 (contours) overlayed on the HST image of the ``dumbbell'' galaxy
 (grayscale). Note that both nuclei show up in the radio image, albeit only the
 strongest has powerful jets. At the distance of NGC326, 10\arcsec~corresponds
 to 9.3 kpc ($H_{0}=75~{\rm km}~{\rm s}^{-1}~{\rm Mpc}^{-1}$).}
\end{figure}
We present VLA observations of the radio galaxy NGC326 at wavelengths of 
 2, 3.5, 6, and 20 cm. Data from different arrays were combined in order to
 improve uv-coverage and sensitivity. Fig.~1 shows four images of the radio
 galaxy in order of increasing resolution.
 At a resolution of 10\arcsec~the most prominent components appear to be
 the lobe and the two wings. As already noted by  Worrall et al. (1995),
 the overall `Z'-shaped symmetry of the source is broken by a low surface
 brightness plume located just above the end of the north-east wing.
 This plume, evident only in the 20 and 6 cm images, 
appears not to obey the source symmetry and this constitutes a problem
 for the jet precession/realignment scenarios.
 The 15~GHz, 4\arcsec~resolution, image shows that the north-west jet is,
 in projection, shorter with respect to the south-east jet and
 the position of the core does not coincide with the center of symmetry 
of the extended structure. NGC326 is a double system composed of two nearly 
equally bright elliptical galaxies in a common envelope (``dumbbell'' galaxy).
 Fig.~1-D shows the VLA image of NGC326 overlayed on the HST image
 of the galaxy.
\section{Spectral analysis}
 For the purpose of the spectral and polarization analysis we have
 made data sets composed of matched resolution images. These images
 allow us to trace in detail the spectral trends in all regions 
of the source (see Fig.~2).
 Further investigations of these spectral trends will give  
 important clues about the interpretation of this peculiar radio galaxy.
\acknowledgments
We are grateful to Larry Rudnick for patiently explaining us the tomography 
 technique and for stimulating discussions.
 MM acknowledge a partial support by the Italian Ministry for University and Research (MURST) under grant Cofin98-02-32.
\begin{figure}
\plotfiddle{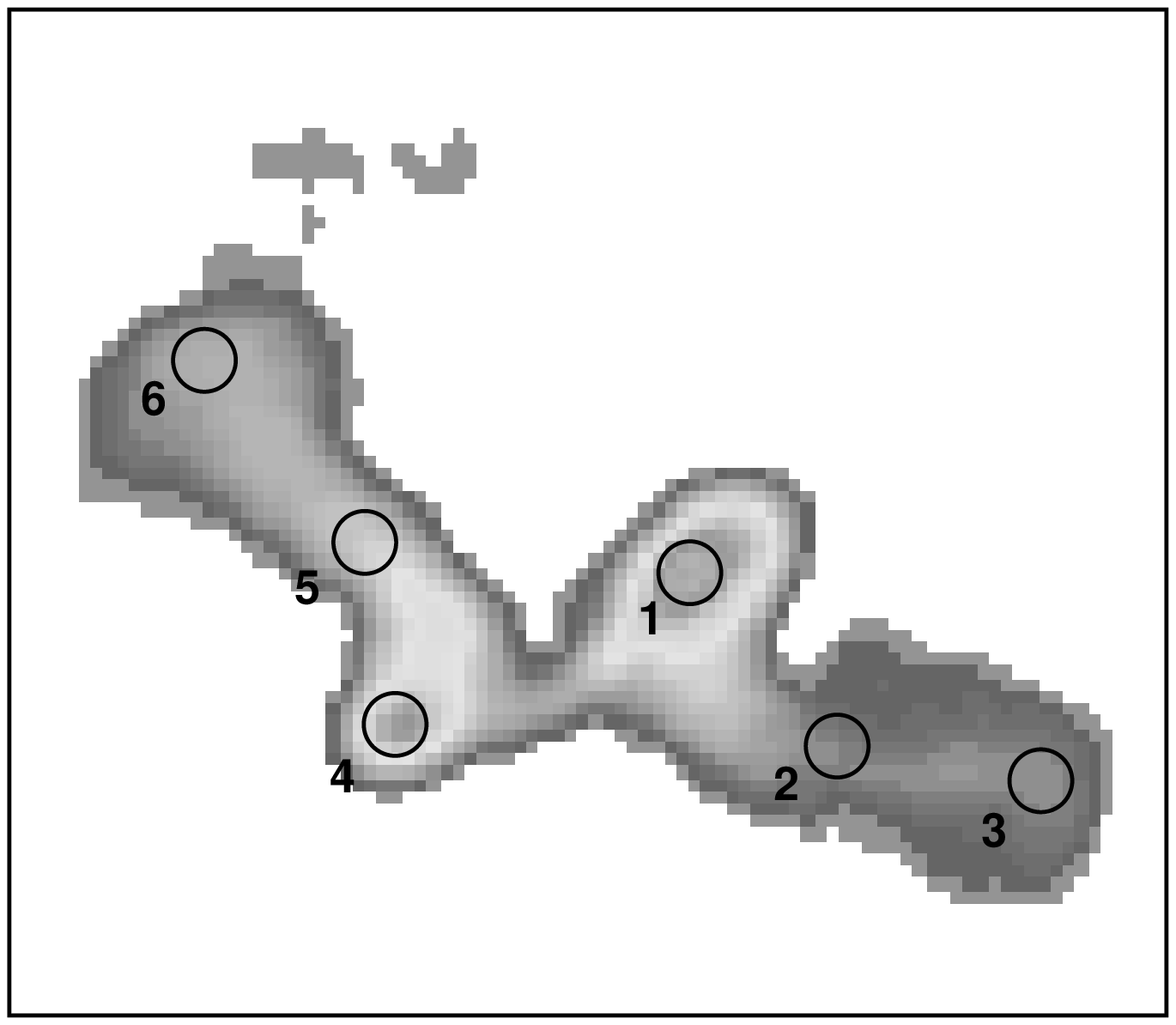}{8cm}{0.0}{40}{40}{-125.0}{50.0} 	
\plotfiddle{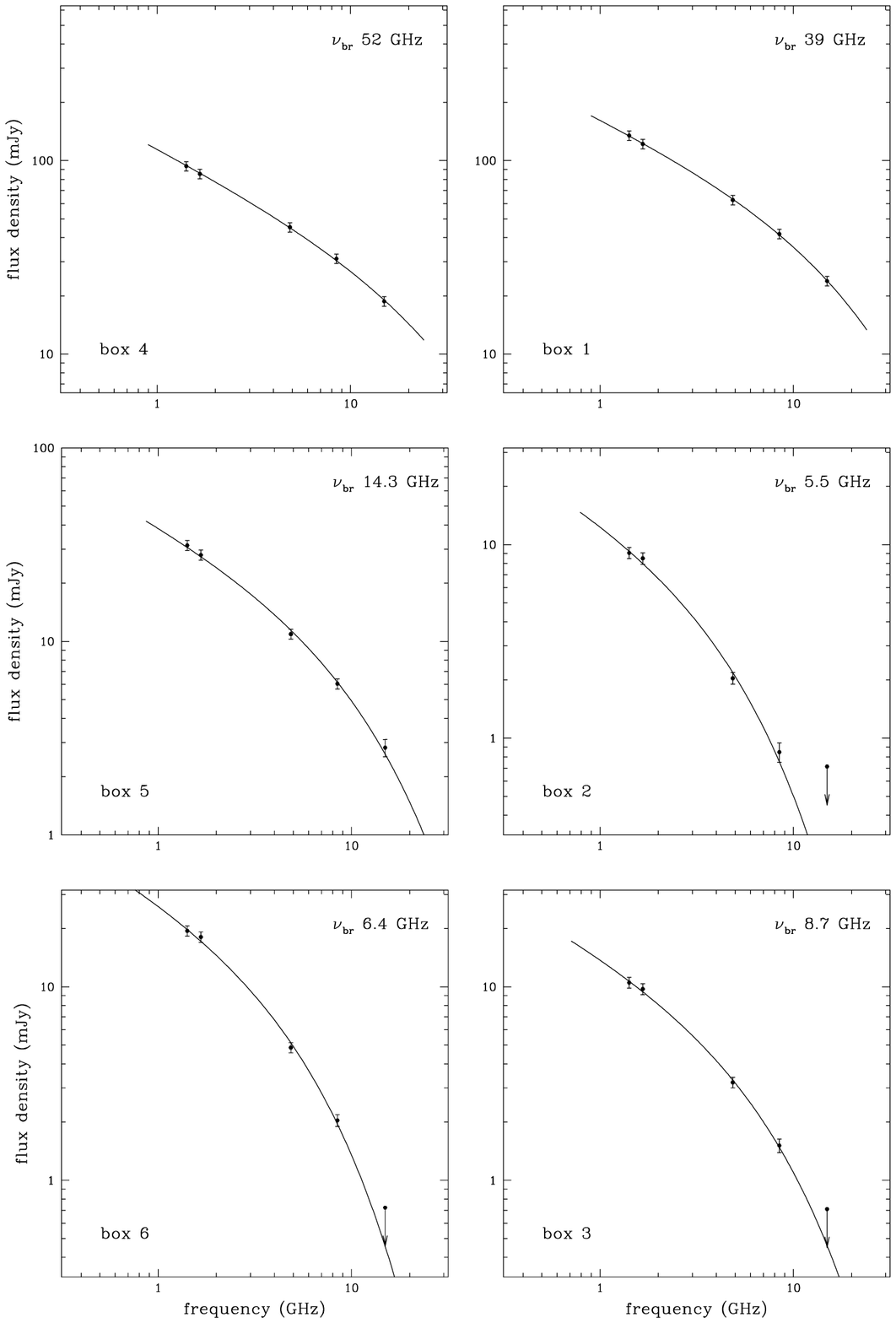}{10cm}{0.0}{65}{65}{-200.0}{-25.0} 	
\caption{Spectra taken at various positions along the source. The solid 
line represents the best fit of a standard JP model (Jaffe \& Perola 1973) 
 with a fixed injection spectral index of $\alpha_{\rm inj}=0.5$.
 The north-east wing (left panels) shows a clear and  monotonic decrease of
 the spectral break ($\nu_{\rm br}$) as a function
 of the distance from the south-east lobe. The south-west wing (right panels) 
shows a different and complex spectral behavior.
 After an initial decrease, $\nu_{\rm br}$ starts to move to high frequency 
going towards the end of the wing.}
\end{figure}


\begin{references}
Ekers, R. D., Fanti, R., Lari, C., \& Parma, P. 1978, `NGC326 - A radio galaxy with a precessing beam', Nature 276, 588-590\\
Jaffe W.J., Perola G.C. `Dynamical Models of Tailed Radio Sources in Clusters of Galaxies', 1974, A\&A 26, 423-435\\
Rees, M. J. 1978, `Relativistic jets and beams in radio galaxies', Nature 275, 516-517\\
Wirth, A., Smarr, L.,\& Gallagher, J. S. 1982, `Dumbbell galaxies and precessing radio jets', AJ 87, 602-615\\
Worrall, D. M., Birkinshaw, M.,\& Camero, R. A. 1995, `The X-Ray Environment of the Dumbbell Radio Galaxy NGC 326', ApJ 449, 93-104
\end{references}
\end{document}